\documentstyle[12pt]{article}
\begin{document}

\title{Diffusion in disordered\\
systems under iterative measurement}
\author{J. C. Flores}
\date{Universidad de Tarapac\'a, Departamento de F\'\i sica, \\
Casilla 7-D, Arica-Chile\\
}
\maketitle

\baselineskip=14pt

We consider a sequence of idealized measurements of time-separation $\Delta
t $ onto a discrete one-dimensional disordered system. A connection with
Markov chains is found. For a rapid sequence of measurements, a diffusive
regime occurs and the diffusion coefficient $D$ is analytically calculated.
In a general point of view, this result suggests the possibility to break
the Anderson localization due to decoherence effects. Quantum Zeno effect
emerges because the diffusion coefficient $D$ vanishes at the limit $\Delta
t\rightarrow 0$.

$$
{} 
$$

$$
{} 
$$


Accepted in Phys.Rev.B

\newpage 
$$
{} 
$$

In this work we study the effect of ideal successive measurements onto a
disordered one-dimensional random system. The iterative measurement,
separated by intervals of time $\Delta t$, are realized on the position
observable and they are responsible for a diffusive regime in the
approximation $\Delta tb/\hbar \ll 1$. In the general case (i.e. for any $%
\Delta t$) the problem becomes related to a classical Markov chain equations
for the probability, not the amplitude. The diffusive regime is related to
coherence destruction, operating from the macroscopic measurement apparatus.
Since Anderson localization [1-5] operates on coherent systems like
Schr\"odinger equation, we can expect a diffusive behavior for open systems.
In fact, for coherent disordered systems, the return-probability at the
initial position is nonzero when $t\rightarrow +\infty $ (periodical motion
[3,4]) and opposite to the usual nonreversible dynamics behavior expected
for open systems.

$$
{} 
$$

Explicitly, consider a random one-dimensional tight-binding model with
nearest neighbor hopping Hamiltonian (Anderson model),

\begin{equation}
H=\sum_l\xi _l\mid l\rangle \langle l\mid +b\{\mid l\rangle \langle l+1\mid
+\mid l+1\rangle \langle l\mid \} 
\end{equation}
where the integer $l$ denotes the lattice position, $\xi _l$ are random
independent variables, $b$ is the hopping constant and $\mid l$ $\rangle $
stands for the discrete position basis. Moreover, the position operator is
given by its spectral decomposition

\begin{equation}
Q=a\sum_ll\mid l\rangle \langle l\mid . 
\end{equation}
where the length $a$ denotes the period of the lattice. It is well known
that the wave function becomes exponentially localized for a disordered
system like in equation (1). Moreover, from Landauer formula for the
resistance, no conduction is expected [5].

$$
{} 
$$

Consider a set of $N+1$ idealized measures on the observable position $Q$ of
the system. The measures are separated by bounded intervals of time $\Delta
t_n$ ($n=1,2,...N$); which can be constant, random, or others. Let $\rho (t)$
be the statistical operator describing the disordered system. Using the von
Neumann measure scheme [6,7], and since the Hamiltonian $H$ does not commute
with the position operator $Q$, the mapping between two consecutive measures
is given by

\begin{equation}
\rho _{n+1}^{(+)}=\sum_lP_l\left( e^{iH\Delta t_n/\hbar }\rho
_n^{(+)}e^{-iH\Delta t_n/\hbar }\right) P_l 
\end{equation}
where $\rho _n^{(+)}$ stands for the statistical operator just after the
measurement $n,$ and $P_l$ is the corresponding projector operator onto the
sub-space spanned by the vector $\mid l\rangle $. Namely,

\begin{equation}
P_l=\mid l\rangle \langle l\mid . 
\end{equation}

In this way, we have a free evolution ( $t_n^{+}$ $\rightarrow t_{n+1}^{-}$)
carried out with the usual quantum mechanics unitary operator $U=e^{iH\Delta
t/\hbar }$. And at the instant $t_{n+1}$ operates the measurement process.
After the measurement, the statistical operator becomes diagonal in the
representation of the observable [6,7]. Thus, necessarily the density
operator can be written as

\begin{equation}
\rho _n^{(+)}=\sum_lW_l^{(n)}P_l 
\end{equation}
where $W_l^{(n)}$ is the probability to find the particle at position $l$
and `time' $n$. From (3) and (5), the evolution equation for the probability
becomes,

\begin{equation}
W_l^{(n+1)}=\sum_s\parallel \langle l\mid e^{iH\Delta t_n/\hbar }\mid
s\rangle \parallel ^2W_s^{(n)} 
\end{equation}
where, for the transition matrix, each row adds up to unity and $%
\sum_lW_l^{(n)}=1$ corresponding to a Markov-chain. Moreover, it corresponds
to a doubly stochastic transition matrix i.e. not only the row sums but also
the column sums are unity.

$$
{} 
$$

For a finite system, i.e. $-L\leq l\leq L$, the unique stationary
probability solution is $W_l^{(\infty )}=\frac 1{2L+1}$ corresponding to an
extended state. To show that $W^\infty $ is the unique solution we operate
through contradiction: assume other solution $G^\infty $, then necessarily
it is not orthogonal to $W^\infty $ because $\sum_lW_l^\infty G_l^\infty
=\frac 1{2L+1}$. This vector can be decomposed as $G^\infty =\alpha G^{\perp
}+\beta W^\infty $ where $\alpha $ and $\beta \neq 0$ are constants and $%
G^{\perp }$ is a vector in the orthogonal space of $W^\infty $. But since
(6) is a linear equation, $G^{\perp }$ is also a stationary solution, in
contradiction with the no existence of orthogonal solutions.

$$
{} 
$$

In this way, an iterative measurement process breaks the Anderson
localization. The origin of this delocalization is very different to others,
related to internal correlations [8-16] or special random distributions
[17,18], where the evolution is always represented as a pure state
(Schr\"odinger equation). In our case the measurement process produces
mixture and diffusion from decoherence effects.

$$
{} 
$$

Calculations can be explicitly made if we consider the approximation of
small intervals of time, that is when $\Delta t_nb/\hbar \ll 1$. Using the
well known expansion

\begin{equation}
e^{iH\Delta t/\hbar }\rho e^{-iH\Delta t/\hbar }=\rho +i\frac 1\hbar \Delta
t[\rho ,H]-\frac 1{2\hbar ^2}\left( \Delta t\right) ^2[H,[H,\rho ]]+O(\Delta
t^3), 
\end{equation}
the first order expansion in equation (3), gives us the master evolution
equation for the probability

\begin{equation}
W_l^{(n+1)}-W_l^{(n)}=\frac 1{\hbar ^2}\left( \Delta t_n\right)
^2\sum_s\left\| \langle l\mid H\mid s\rangle \right\| ^2\left(
W_s^{(n)}-W_l^{(n)}\right) , 
\end{equation}
namely, quadratic in the expansion parameter. Defining the spatial
dispersion in the usual way $\sigma _n^2=a^2\sum_ll^2W_l^{(n)}$. Then, from
equation (8) we have the discrete evolution equation

\begin{equation}
\sigma _{n+1}^2-\sigma _n^2=\frac{2b^2a^2}{\hbar ^2}\left( \Delta t_n\right)
^2, 
\end{equation}
and defined only within a small time interval. In this equation, different
possibilities must be considered:

$$
{} 
$$

(i) If the interval of time between two consecutive measurements is constant
($\Delta t_n=\tau )$, the dispersion has a diffusive behavior given by

\begin{equation}
\sigma _n^2=\frac{2b^2a^2}{\hbar ^2}\left( \tau \right) ^2n, 
\end{equation}
where we have assumed the initial dispersion $\sigma _0=0$. From equation
(10), the diffusion coefficient $D$ is ($t=n\tau $)

\begin{equation}
D=\frac{2b^2a^2\tau ^{}}{\hbar ^2}. 
\end{equation}

(ii) If the intervals $\Delta t_n$are random bounded independent quantities.
Also a diffusive regime must be expected because the average $\langle \Delta
t_n\rangle $ are independent of $n$. Remark that from (9) the dispersion $%
\sigma _{n\rightarrow \infty }^2$ becomes Gaussian distributed (central
limit theorem).

$$
{} 
$$

(iii) The intervals $\Delta t_n$ are constant and corresponding to $N$
measurement realized in the total interval of time $T$, namely $\tau =\frac
TN$ in (10) and (11). Here the diffusion coefficient becomes

\begin{equation}
D=\frac{2b^2a^2T^{}}{\hbar ^2N} 
\end{equation}
and vanishes in the limit $N\rightarrow \infty $ (quantum Zeno effect [19]).
Namely, at this limit the system remains freezed at the first state after
the measurement.

$$
{} 
$$

In conclusion we have presented theoretical evidence that idealized
iterative measurement break strong localization. This result confirms that
decoherence effects produce diffusion in disordered systems. This is a
general result related to the absence of a discrete spectrum for open
systems. To be more explicit, consider the linear entropy $S(t)$ defined as

\begin{equation}
S=1-Tr\rho ^2, 
\end{equation}
then the measurement process increases $S$ irreversibly because $%
S^{(+)}>S^{(-)}.$ The spectrum cannot be discrete, and the particle does not
reach its initial position as usually in Anderson localization. In other
more general disordered open systems, we can expect the relationship

\begin{equation}
\frac{dS}{dt}>0 
\end{equation}
linked to irreversibility. As said before, for disordered systems, the
probability of return to the initial position is not zero (periodical motion
due to discrete spectrum) and it is incompatible with (14). 
$$
{} 
$$

Finally we remark that recently it was shown that the ortho-para nuclear
spin conversion in CH$_3$F molecules relaxes having a rate which decreases
when the gas pressure is increasing [20]. This is an example of freezing of
quantum states induced by collisions, and is an experimental support to the
quantum Zeno effect. In our case, the inhibition effect is present in the
appropriate limit (12). Nevertheless, the interesting fact here is that in
the spin conversion experiment the monitoring is carried out using the
surrounding gas. Recently [21] the breaking of the Anderson localization in
systems with correlated disorder has been experimentally found using
semiconductor superlattices (SL's), by means of optical and transport
measurements. Thus we are thinking on suitable SL's, in order to verify the
theory of the present work. We are considering ultrafast optical
spectroscopy, using femto-second laser pulses, as experimental technique
that has been used recently to measure experimentally the Bloch electron
displacement as a function of time in SL's [22].

$$
{} 
$$

{\bf Acknowledge} : I thanks Dr. V. Bellani and Dr. E. Diez for its comment
about the manuscript and possible experimental realizations on SL's. This
work was motivated by discussions at the `International Workshop On
Disordered Systems With Correlated-Disorder' (PELICAN Project, Arica, Sept.
98). A preliminary version of this work was first presented at the Fourth
International Andino Meeting of Physics (PELICAN Project, Arica, Oct.98).
This work was not supported by FONDECYT.

\newpage\

\end{document}